# Influence of carbon-ion irradiation on the superconducting critical properties of MgB$_2$ thin films

**Soon-Gil Jung[1,2], Seung-Ku Son[1,2], Duong Pham[2], Weon Cheol Lim[3], Jonghan Song[3], Won Nam Kang[2], and Tuson Park[1,2]**

[1]Center for Quantum Materials and Superconductivity (CQMS), Sungkyunkwan University, Suwon 16419, Republic of Korea

[2]Department of Physics, Sungkyunkwan University, Suwon 16419, Republic of Korea

[3]Advanced Analysis Center, Korea Institute of Science and Technology (KIST), Hwarang-ro 14-gil 5, Seongbuk-gu, Seoul 02792, Republic of Korea

E-mail: wnkang@skku.edu and tp8701@skku.edu

Fax: +82 31 290 7056

**Abstract**

We investigate the influence of carbon-ion irradiation on the superconducting critical properties of MgB$_2$ thin films. MgB$_2$ films of two thicknesses viz. 400 nm (MB400nm) and 800 nm (MB800nm) were irradiated by 350 keV C ions having a wide range of fluence, $1 \times 10^{13} - 1 \times 10^{15}$ C atoms/cm$^2$. The mean projected range ($R_p$) of 350 keV C ions in MgB$_2$ is 560 nm, thus the energetic C ions will pass through the MB400nm, whereas the ions will remain into the MB800nm. The superconducting transition temperature

($T_c$), upper critical field ($H_{c2}$), *c*-axis lattice parameter, and corrected residual resistivity ($\rho_{corr}$) of both the films showed similar trends with the variation of fluence. However, a disparate behavior in the superconducting phase transition was observed in the MB800nm when the fluence was larger than $1\times10^{14}$ C atoms/cm$^2$ because of the different $T_c$s between the irradiated and non-irradiated parts of the film. Interestingly, the superconducting critical properties, such as $T_c$, $H_{c2}$, and $J_c$, of the irradiated MgB$_2$ films, as well as the lattice parameter, were almost restored to those in the pristine state after a thermal annealing procedure. These results demonstrate that the atomic lattice distortion induced by C-ion irradiation is the main reason for the change in the superconducting properties of MgB$_2$ films.

## 1. Introduction

Magnesium diboride (MgB$_2$) has promising superconducting (SC) properties for technological applications, such as a high transition temperature ($T_c$) of ~ 40 K [1], high upper critical field ($H_{c2}$) [2,3], and large critical current density ($J_c$) [4,5]. In addition, two clear SC gaps of $\Delta_\sigma$ and $\Delta_\pi$ in MgB$_2$ have stimulated studies to understand the mechanism of its superconductivity [2,6,7]. Carbon (C) substitution for boron (B), MgB$_{2-x}$C$_x$, is well known as one of the most effective methods to improve $H_{c2}$ and $J_c$ despite a reduction in $T_c$ caused by a suppression of the large SC gap $\Delta_\sigma$. The C substitution considerably affects the $\sigma$-bonding derived from the in-plane $p_{xy}$ B orbitals, whereas it hardly changes the $\pi$-bonding related to the out-of-plane $p_z$ B orbitals [2,8-11].

Neutron irradiation in MgB$_2$ also has a significant influence on the B sites because of the thermal neutron capture reaction. The $^{10}$B atoms (~ 20% in natural B) that capture the thermal neutrons decay into an alpha ($\alpha$) particle and a $^7$Li nucleus, leading to homogeneous disorders in MgB$_2$ and an improvement in both $H_{c2}$ and $J_c$ [2,12-15]. In addition, light particles, such as protons and $\alpha$ particles, have often been reported to be effective sources of irradiation to enhance $H_{c2}$ and $J_c$ of MgB$_2$ [16,17]. On the other hand, heavy-ion irradiations are less effective in improving $J_c$ of MgB$_2$ [18,19]. In general, high energy heavy-ion irradiations into SC materials produce columnar defects along the ion tracks, which act as a strong flux-pinning source to improve the in-field $J_c$ of the high-$T_c$ cuprate superconductors [20,21]. However, heavy-ion irradiations in MgB$_2$ seem to rapidly suppress its superconductivity rather than improve its critical SC parameters, such as $H_{c2}$ and $J_c$ [18,22]. For these reasons, irradiation studies of $MgB_2$ have been mostly focused on neutron irradiation.

Irradiation is a unique technique that makes it possible to examine the effects of radiation and additional defects in an identical sample [13,23,24]. The study of radiation effects on a superconductor is important to test their application in a radiation environment, such as space, fusion reactions, and so on [13,23]. In addition, ion irradiations can create defects in target materials mainly by nuclear stopping of energetic ions, although the exact mechanism for the formation of defects is still debatable [23,25]. In superconductors, defects are normally beneficial for the enhancement of SC critical properties, especially in-field $J_c$, because defects in type-II superconductors can interrupt vortex motion. Therefore, understanding the effects of defects on SC critical properties is a key issue for technological applications of superconductors as well as for scientific advancement [26,27].

In this study, we report the influence of carbon-ion irradiation on the superconducting critical properties of *c*-axis-oriented $MgB_2$ thin films with thicknesses of 400 and 800 nm. The C was selected as the irradiating ion source because C has been known as one of the most effective elements to enhance $H_{c2}$ and in-field $J_c$ of $MgB_2$. Carbon ions of various fluences ranging between $1 \times 10^{13}$ and $1 \times 10^{15}$ C atoms/cm$^2$ with an incident energy of 350 keV (mean projected range $R_p$: ~ 560 nm) were irradiated into the prepared $MgB_2$ films. The superconducting properties after irradiations were slightly different depending on whether the irradiated C ions penetrated fully the film (400 nm) or were implanted into the film (800 nm). However, the degraded superconductivity of both the irradiated $MgB_2$ films was almost recovered to those in the pristine state after a thermal annealing procedure. These results indicate that the effect of the implanted C ions on the superconductivity of $MgB_2$ films is insignificant, whereas the displacement damage by irradiation is crucial to the change in $T_c$, $H_{c2}$, and $J_c$ of $MgB_2$.

## 2. Experimental

High-quality $MgB_2$ thin films of thicknesses 400 nm (MB400nm) and 800 nm (MB800nm) were fabricated on *c*-cut $Al_2O_3$ substrates for C-ion irradiation studies by using a hybrid physical-chemical vapor deposition (HPCVD) method, the details of the growth technique described earlier [28,29]. C-ion irradiation was carried out using Cockcroft-Walton type 400 kV ion beam accelerator at Korea Institute of Science and Technology (KIST), Seoul. Various fluences of $1 \times 10^{13}$ (1E13), $5 \times 10^{13}$ (5E13), $1 \times 10^{14}$ (1E14), $2 \times 10^{14}$ (2E14), $5 \times 10^{14}$ (5E14), and $1 \times 10^{15}$ (1E15) C atoms/cm$^2$ with a tilted angle of 7° to avoid channeling effects during irradiations were irradiated into the prepared $MgB_2$ films at room temperature. The mean projected range ($R_p$) of the irradiated C ions, with an incident beam energy of 350 keV, was estimated to be around 560 nm from SRIM (The Stopping and Range of Ions in Matter), a Monte Carlo simulation program [30]. Therefore, the irradiated C ions were remained into the MB800nm due to the $R_p$ being greater than the film thickness whereas most of the ions penetrated through the MB400nm.

The thickness of the $MgB_2$ thin films was measured by using a scanning electron microscope (SEM), and the crystallinity of the films before and after irradiation was checked by using an x-ray diffractometer (XRD). The electrical resistivity ($\rho$) was measured by using the standard 4-probe technique with gold (Au) coating on the 4-point contact regions to achieve good ohmic contact. The upper critical field ($H_{c2}$) for a magnetic field applied perpendicular to the *ab* plane of the surface of the film was measured by the Physical Property Measurement System (PPMS 9 T, Quantum Design) and the temperature dependence of magnetization, $M(T)$, was measured by using the Magnetic Property Measurement System (MPMS 5 T, Quantum Design) before and after C-ion irradiation.

Thermal annealing was carried out for the MB400nm and MB800nm irradiated with the highest fluence of $1 \times 10^{15}$ atoms/cm$^2$ (1E15). In addition, a non-irradiated MB400nm was also prepared for the thermal annealing to check only annealing effect on the pristine sample. Three films were wrapped in tantalum (Ta) foil to minimize the Mg decomposition during thermal annealing, sealed inside an evacuated quartz tube, and then the ampoules were annealed at temperatures of 200, 300, 400, and 500 °C for 30 min. in a box furnace. The same irradiated films with the 1E15 were consecutively annealed at each temperature, and the non-irradiated MB400nm was annealed only at 500 °C. The annealing experiments were conducted at temperatures up to 500 °C as large Mg deficiency can be induced at higher temperatures [31]. $M(T)$ and x-ray diffraction patterns of the irradiated films were examined after the annealing procedure for each annealing temperature ($T_A$). In addition, $\rho$, $H_{c2}$, and magnetization hysteresis ($M - H$) loops for the films annealed at $T_A = 500$ °C were investigated by using the same techniques mentioned above.

## 3. Results and discussion

Figure 1 describes the mean projected range ($R_p$) of the irradiated C ions with a beam energy of 350 keV into MgB$_2$, simulated by using SRIM (Stopping and Range of Ions in Matter) software [30], presenting that most of the C ions are stopped at around $R_p \sim 560$ nm after undergoing a series of collisions with the Mg and B atoms. In view of the magnitude of the $R_p$, two MgB$_2$ thin films having thicknesses of 400 nm (MB400nm) and 800 nm (MB800nm) were irradiated with the 350 keV C ions, and while most of the energetic C ions penetrated through MB400nm, as shown in figure 1, they remained in MB800nm as defects or were substituted into B sites.

Figures 2(a) and (b) show the temperature dependence of the normalized electrical resistivity, $\rho(T)/\rho(41\ K)$, near the superconducting (SC) transition regions for pristine (pri.) and irradiated MgB$_2$ films of MB400nm and MB800nm, respectively, whereby C-ion irradiation into MgB$_2$ thin films resulted in suppression of the SC transition temperature ($T_c$). In this case, $T_c$ was determined from 50% of the normal state resistivity value at SC onset temperature, as indicated by arrows in figures 2(a) and (b). For the films irradiated with higher fluences than $1 \times 10^{14}$ atoms/cm$^2$ (1E14), the SC transition width of MB800nm is much broader than that of MB400nm, while $T_c$ of MB800nm is higher than that of MB400nm. As shown in figures 2(c) and (d), the zero-field-cooled (ZFC) and field-cooled (FC) dc magnetization ($M$) data for MB400nm and MB800nm also showed similar results to the electrical resistivity. Here, $M(T)$ values were normalized by the absolute ZFC $M$ value at 5 K for comparison. The onset temperature for the diamagnetic signal gradually reduces with increasing fluence for both MB400nm and MB800nm. A kink in the ZFC $M(T)$ data, however, is apparent in the SC transition region for MB800nm irradiated with large fluences of $5 \times 10^{14}$ (5E14) and $1 \times 10^{15}$ atoms/cm$^2$ (1E15), while it is absent for MB400nm. Since the thickness of MB800nm is larger than $R_p$ (~ 560 nm), the kink observed in the MB800nm may be ascribed to two different $T_c$s between the irradiated and non-irradiated parts of the MgB$_2$ films.

The temperature dependence of electrical resistivity ($\rho$) is comparatively plotted in figures 3(a) and (b) for MB400nm and MB800nm before and after irradiation, respectively. The resistivity increases over the whole temperature range with increasing fluence of the C-ion radiation. Figures 3(c) and (d) present a plot of $T_c$ as a function of the residual resistivity, $\rho_{41K} = \rho(41\ K)$, and corrected residual resistivity ($\rho_{corr}$) for MB400nm and MB800nm, respectively, where $\rho_{corr}$ was introduced to exclude the effects of grain boundaries on the resistivity and is defined as $\rho_{corr} = \rho_{41K} \times \Delta\rho_{ideal}/\Delta\rho$, where $\Delta\rho = \rho(300\ K) - \rho(41\ K)$ and $\Delta\rho_{ideal}$ is $\Delta\rho$ of $MgB_2$ with full connectivity [12,15,32]. Here, we used $\Delta\rho_{ideal} = 4.3\ \mu\Omega\cdot cm$ determined from a $MgB_2$ single crystal because $\Delta\rho$ of the pristine MB400nm and MB800nm is 6.4 $\mu\Omega\cdot cm$ and 5.9 $\mu\Omega\cdot cm$, respectively [32]. For fluences higher than 1E14, the $\rho_{41K}$ values of MB800nm is much larger than those of MB400nm, while the reduction rate of $T_c$ for MB800nm is weaker than that for MB400nm. When $\rho_{41K}$ is converted to $\rho_{corr}$, however, $T_c$s of MB400nm and MB800nm are well scaled on a single curve, as shown in figure 3(d) [13,15]. These results indicate that, despite a larger residual resistivity than MB400nm, the higher $T_c$ in MB800nm results from the proximity effect between the irradiated and non-irradiated parts in it. In addition, the decrease in $T_c$ of both the films can be attributed to enhanced intra-grain scattering by C-ion irradiation [13,32], whereas the effect of implanted C ions on $T_c$ seems negligible in the MB800nm films.

Figures 4(a) and (b) show the x-ray diffraction (XRD) patterns of a $\theta – 2\theta$ scan for pristine and irradiated MB400nm and MB800nm, respectively, showing a shift of (00$l$) peak positions of $MgB_2$ due to C-ion irradiation. Figures 4(c) and (d) are the magnified views of the (002) peak of MB400nm and MB800nm, respectively. With increasing fluence, the peak position is progressively shifted towards a lower angle and the peak width is broadened [12,14,15,33], reflecting an expansion of the *c*-axis lattice parameter and degradation of crystallinity of the MgB$_2$ films. The (002) peak is considerably broadened for the irradiated MB800nm than for the MB400nm. In addition, MB800nm irradiated with 1E15 show a double-peak like behavior due to the implanted C ions around the $R_p$, as shown in figure 1.

The dependence on C-ion fluence of the *c*-axis lattice constant, $\rho_{41K}$, and $T_c$ for MB400nm and MB800nm are plotted in figures 5(a), (b), and (c), respectively. In comparison with the pristine films, the *c*-axis lattice parameter of MB400nm and MB800nm irradiated with 1E15 fluence is increased by 0.75% and 1.22%, respectively. The expanded lattice parameter in irradiated films underlines the fact that displacement damage by nuclear stopping (elastic scattering between energetic C ions and Mg/B atoms) is more prominent than the damage by electronic stopping (interaction between the energetic C ions and electrons of MgB$_2$) for an increase of $\rho_{41K}$ and suppression of $T_c$ [23]. A sudden increase in $\rho_{41K}$ of MB800nm at $2\times10^{14}$ atoms/cm$^2$ (2E14) implies that the implanted C ions create substantial defects due to large damage events around $R_p$, while the $T_c$ suppression is smeared out due to the proximity effects between the irradiated and non-irradiated parts of the MgB$_2$ films.

In order to study the effect of C-ion irradiation on the upper critical field ($H_{c2}$) of MB400nm and MB800nm, the electrical resistivity was measured as a function of temperature for magnetic fields applied perpendicular to the *ab* plane. Figures 6(a) and (b) display the $\rho(T)$ curves in a magnetic field for the pristine and C-ion irradiated MB400nm with 2E14 fluence, respectively. It will be shown that 2E14 is an optimal fluence for the enhancement of $H_{c2}$ in this study. Here, $\rho(T)$ is normalized by the normal state resistivity ($\rho_n$) at each magnetic field for comparison, and $H_{c2}$ is determined as the middle point of the SC transition, as indicated by arrows in figures 6(a) and (b). $H_{c2}$s of MB400nm and MB800nm plotted as a function of temperature for various fluences are shown in figures 6(c) and 6 (d), respectively, where all $H_{c2}(T)$ curves are well explained by the two-band Ginzburg-Landua (GL) theory [34]. The solid lines in figures 6(c) and (d) are representative fitting curves for $H_{c2}(T)$ of the pristine and MgB$_2$ films irradiated with a fluence of 2E14.

The upper critical field at zero Kelvin, $H_{c2}(0)$, estimated from the data in figures 6(c) and (d) is plotted as a function of fluence and $T_c$ in figures 7(a) and (b), respectively. $H_{c2}(0)$ of the pristine MB400nm and MB800nm is 6.16 and 5.85 T, respectively, which increases rapidly with increasing fluence and reaches a maximum of 11.09 T for MB400nm and 9.78 T for MB800nm at 2E14, and decreases slowly with further increasing fluence. Both the films show the largest enhancement of $H_{c2}(0)$ at the same fluence of 2E14. Even though there occurs a large improvement in $H_{c2}(0)$ by C-ion irradiation, the $T_c$ suppression in MgB$_2$ films by C-ion irradiation, as shown in figure 7(b), is relatively small and comparable with the results of irradiation of MgB$_2$ by other particles, such as neutrons, $\alpha$ particles, oxygen ions, and so on [12,15,17,33]. In addition, the $T_c$ of MB800nm is insensitive to the irradiation compared to that of MB400nm, suggesting that undamaged SC layer is helpful to improve SC critical properties, such as $H_{c2}$ and critical current density ($J_c$), while minimizing $T_c$ reduction.

Figure 8 shows the evolution of XRD patterns around the (002) peak for the pristine and C-ion irradiated MB400nm and MB800nm with the fluence of 1E15, when the irradiated films were subsequently annealed for 30 min. at $T_A$ = 200, 300, 400, and 500 °C. The (002) peak position of the irradiated film, which is shifted towards a lower angle than that for the pristine film, moves to a higher angle with increasing annealing temperature and is close to that of the pristine film for $T_A$ = 500 $^o$C. Figure 9 shows a plot of the *c*-axis lattice parameter of the irradiated MgB$_2$ films as a function of the annealing temperature. Unlike MB400nm, the *c*-axis lattice parameter of MB800nm does not fully regain its value to that in the pristine state after thermal annealing at 500 $^o$C, indicating that the C ions implanted into MB800nm remain in the films as interstitial defects. These results indicate that lattice distortion by Frenkel disorder, such as interstitials and vacancies, is mainly produced by C ions traveling through the MgB$_2$ because thermal annealing for a short period (30 min.) was sufficient enough to restore the expanded lattice parameter of the irradiated MgB$_2$ films to that in the pristine state [13,35,36].

Figures 10(a) and (b) show the temperature dependence of ZFC and FC dc magnetization (*M*) for the annealed MB400nm and MB800nm films, respectively. The SC regions of both the films degraded by irradiation are gradually recovered as the annealing temperature ($T_A$) increases. Furthermore, the kink observed in MB800nm irradiated with a fluence of 1E15 disappears by thermal annealing, underlining that it originated due to the $T_c$ difference between the irradiated and non-irradiated layers in MB800nm, as discussed in figure 2(d). It was noted that the magnetization of MB400nm did not fully recover to that of the pristine state after annealing at 500 $^o$C, even though its *c*-axis lattice parameter was completely restored, as shown in figures 8(a) and 9. The ZFC transition of the pristine MB400nm become slightly broad after annealed at 500 $^o$C, probably due to a decomposition of Mg during the thermal annealing [31], which would be one of the reasons why the $T_c$ of irradiated films does not recover completely after thermal annealing.

The SC transition temperature ($T_c$) and upper critical field ($H_{c2}$) of the irradiated films (1E15), after annealing them at $T_A$ = 500 $^o$C, were determined from electrical resistivity measurements and compared for various fluences of C-ion irradiation. The results were similar to the *M-T* results presented in figure 10, and the $T_c$ of the annealed films is slightly lower than that of the pure films, as shown in figure 11(a). $T_c$s of the annealed films decrease linearly with the corrected residual resistivity ($\rho_{corr}$) which is relatively larger for the annealed films than the pristine films, and may be ascribed to amorphization in some regions or Mg decomposition during the thermal annealing procedure [31]. The fact that $H_{c2}(0)$ of the annealed films is slightly higher than that of the pristine samples, as presented in figure 11(b), may be due to the same reason as in the case of $\rho_{corr}$. The fluence of 1E15 corresponds to a concentration of ~ 8 × $10^{19}$ C atoms/cm$^3$ at $R_p$, which is much lower than the concentration of MgB$_2$, which is ~ 1.15 × $10^{22}$ Mg atoms/cm$^3$ and ~ 2.29 × $10^{22}$ B atoms/cm$^3$ [37,38]. This relatively small fraction of C ions may be one of the possible reasons for the negligible effects of the implanted C ions as impurities in the annealed MB800nm. More detailed studies, such as multiple injections and microscopic investigation, will be needed to make a definitive conclusion on the role of implanted C ions.

Figure 12 shows the magnetic field dependence of critical current density ($J_c$) at 5 K for MB400nm and MB800nm: pristine (squares), 1E15 (circles), and annealed at 500 °C (triangles). The $J_c$ was estimated from the magnetization hysteresis (*M – H*) loops by using Bean's critical state model ($J_c = 30\Delta M/r$) [5], where $\Delta M$ is the height of the *M – H* loops and *r* is the corresponding radius of the total area of the film's surface, and the *M – H* loops were measured in magnetic fields applied perpendicular to the film's plane. A large self-field $J_c$ of pristine films and its rapid drop in magnetic fields indicate a high quality of MB400nm and MB800nm. The field performance of $J_c$ for both the irradiated films is much stronger than that in the pristine state because of additional pinning sites created by C-ion irradiation, whereas a significant reduction of the $J_c$ at zero field is due to the suppression of superconducting regions and $T_c$ reduction. The $J_c$ value of irradiated MB800nm is slightly larger than that of irradiated MB400nm because of the presence of a non-irradiated layer in MB800nm films. Surprisingly, the field performance of $J_c$ for both the irradiated films almost recovered to that of the pristine state after thermal annealing at $T_A = 500\ ^oC$, which is in sharp contrast to the thermal annealing effects on $J_c(H)$ for neutron-irradiated $MgB_2$ polycrystals [14]. A similar field performance of $J_c$ between the annealed MB400nm and MB800nm is consistent with the results on the $H_{c2}(0)$ described in figure 11.

In neutron-irradiated $MgB_2$, SC properties, such as $T_c$, the $c$-axis lattice constant, and $H_{c2}(0)$, are close to the values in the pristine state after annealing at 500 $^oC$, whereas the field performance of $J_c$ is stronger than that of the pristine state. Since $J_c$, the depinning critical current density, is largely influenced by non-SC regions, such as defects, the thermal annealing effects on $J_c(H)$ for neutron-irradiated $MgB_2$ implies that thermodynamically irreversible defects are formed in $MgB_2$ by neutron irradiation [14,39,40]. Considering that the effects of irradiation and thermal neutron capture reactions are complex, however, it is difficult to understand the exact nature of the defects produced by neutron irradiation. On the other hand, the modified SC properties of high-quality $MgB_2$ thin films, such as $T_c$, $H_{c2}$, and the $c$-axis lattice constant, as well as the enhanced $J_c$ by C-ion irradiation were almost fully reversed to the values in the pristine state after thermal annealing for a short period of time. These results indicate that the degraded superconductivity of $MgB_2$ by C-ion irradiations is mainly associated with the atomic lattice distortion caused by displacement damage, which is a reversible deformation, and the recovery of the SC characteristics to the pristine state by thermal annealing could be ascribed to the recovery of its crystallinity.

4. Conclusions

In conclusion, we studied the effects of carbon-ion irradiation on the superconducting critical properties of two $MgB_2$ thin films with different thicknesses: 400 nm and 800 nm. One film was thinner than the mean projected range ($R_p \sim 560$ nm) of irradiated C ions and the other was thicker than $R_p$. The superconducting transition temperatures ($T_c$) of both the films gradually decreased with increasing fluence of the C ions, and were accompanied by an expansion of the $c$-axis lattice constant. In addition, the residual resistivity ($\rho_{41K}$) and upper critical field ($H_{c2}$) considerably increased after C-ion irradiation because of the defects produced by the irradiation. On the other hand, the effect of the implanted C ions on $H_{c2}$ was not prominent compared to other irradiation effects due to the strong interaction between the irradiated and non-irradiated layers in $MgB_2$ films. Interestingly, the degraded superconductivity and expanded lattice parameter of the irradiated $MgB_2$ were almost regained to that in the pristine state after thermal annealing at 500 °C for 30 min. In particular, the improved field performance of the critical current density ($J_c$) by C-ion irradiation was also restored to that in the pristine state, underpinning the fact that the lattice distortion induced by the irradiation is the main origin of the modified superconducting properties of irradiated $MgB_2$ films.

**Acknowledgements**

This work was supported by the National Research Foundation (NRF) of Korea by a grant funded by the Korean Ministry of Science, ICT and Planning (No. 2012R1A3A2048816) and the Basic Science Research Program through the National Research Foundation of Korea (NRF) funded by the Ministry of Education (NRF-2016R1D1A1B03934513 and NRF-2015R1D1A1A01060382). This work is also supported by KIST institutional program (2V06030).

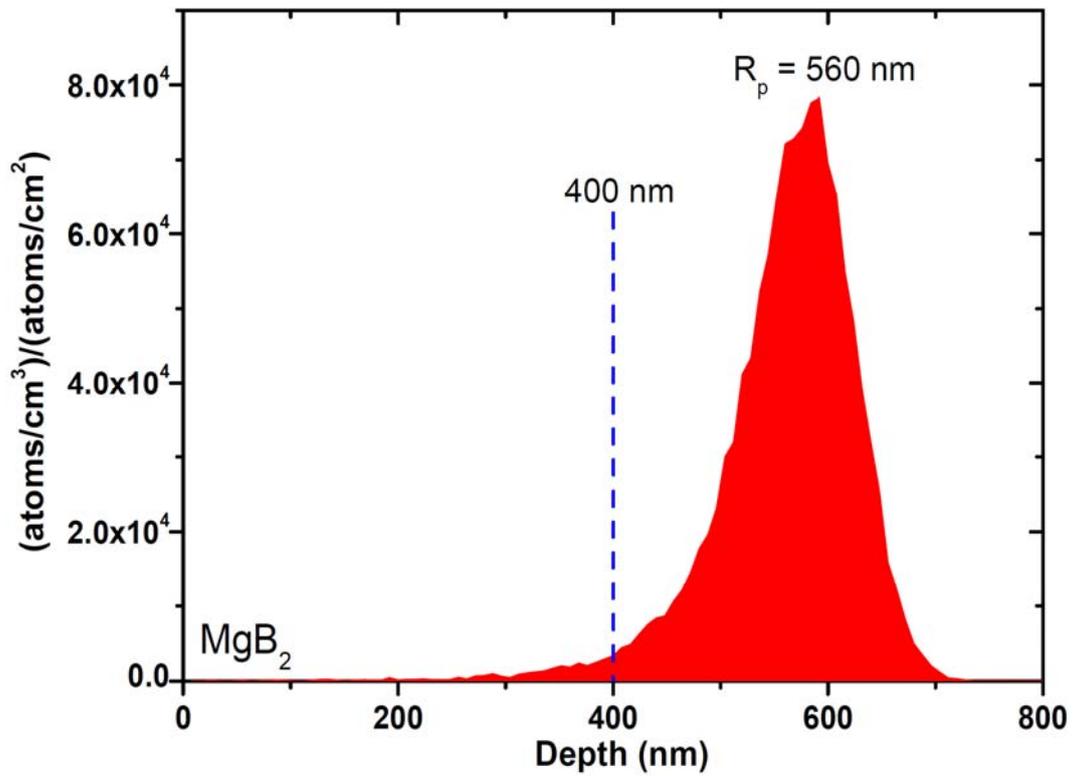

Figure 1. SRIM simulation for the range of irradiated carbon (C) ions with incident energy of 350 keV in $MgB_2$, where the density of $MgB_2$ is 2.57 g/cm$^3$ and the estimated mean projected range ($R_p$) is around 560 nm. Most of the irradiated C ions pass through MB400nm, but for the MB800nm, the ions are distributed around 560 nm below the $MgB_2$ surface.

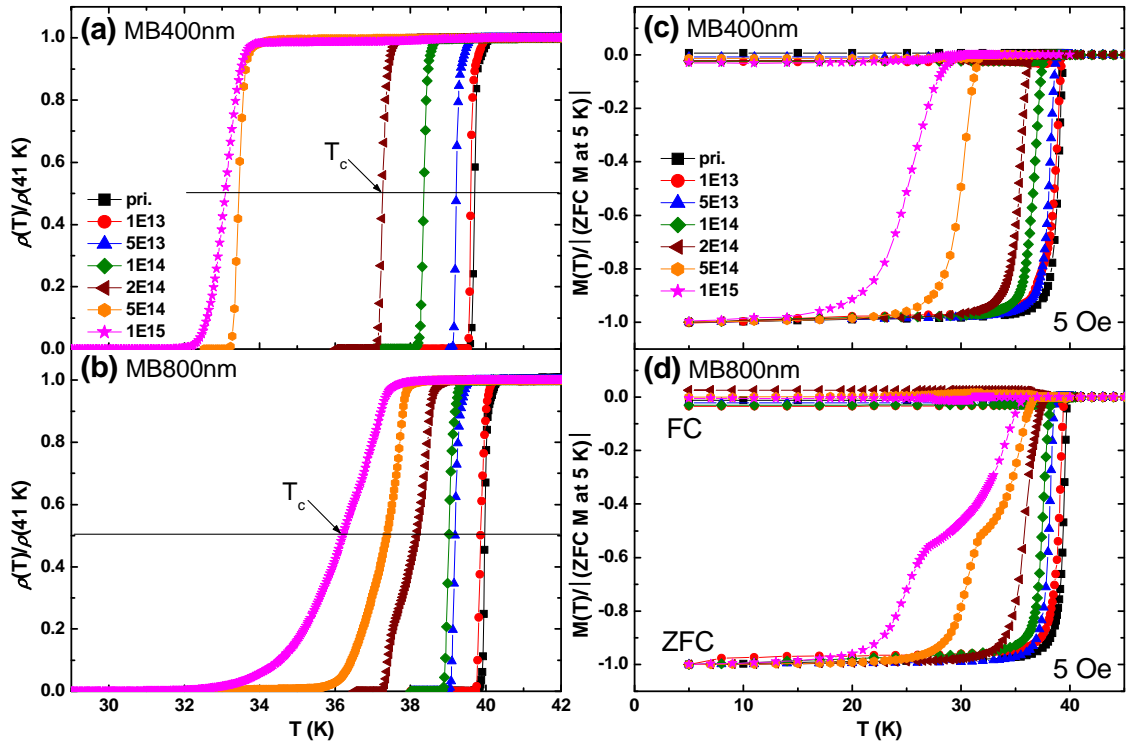

Figure 2. Superconducting transitions measured by [(a) and (b)] electrical resistivity ($\rho$) and [(c) and (d)] magnetization ($M$) for MB400nm and MB800nm at different damage levels of irradiation. Kinks in the $M(T)$ of MB800nm result from a different $T_c$ between the irradiated and non-irradiated layers of $MgB_2$.

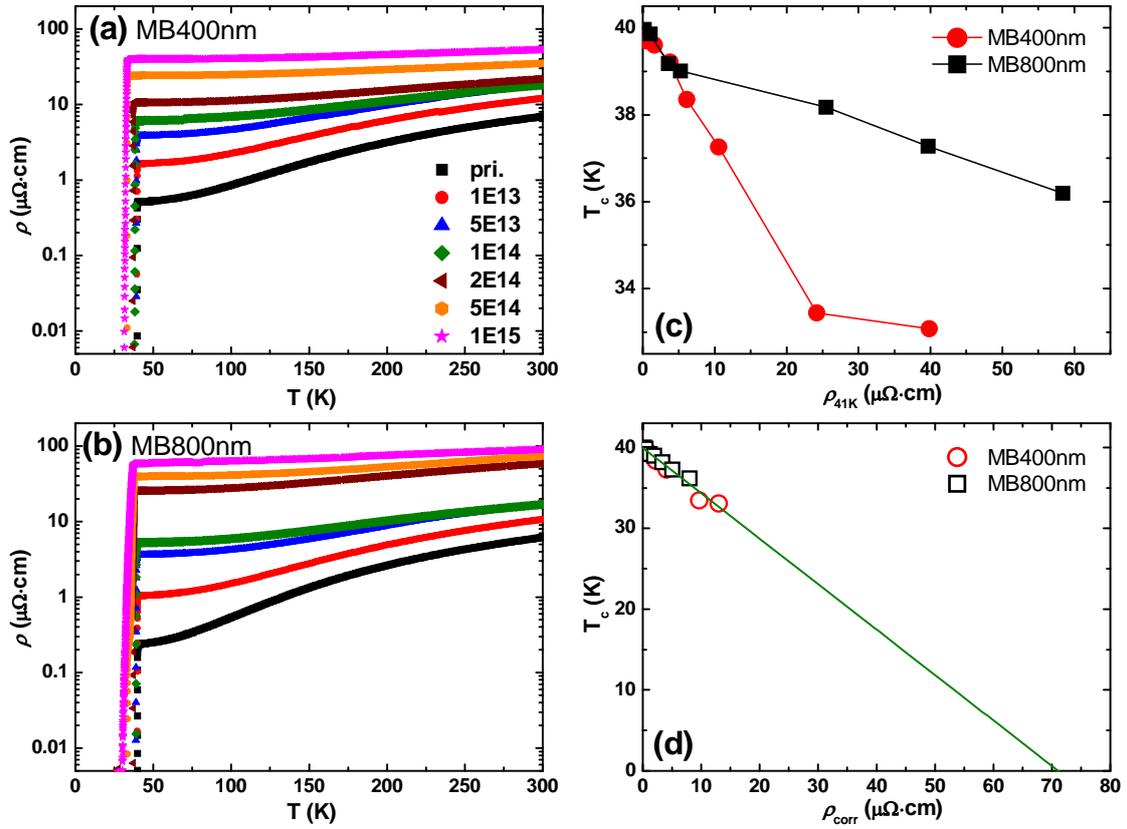

Figure 3. (a) and (b) Temperature dependence of resistivity ($\rho$) of MB400nm and MB800nm, respectively. (c) and (d) Superconducting transition temperature ($T_c$) of MB400nm and MB800nm as functions of residual resistivity at 41 K ($\rho_{41K}$) and corrected residual resistivity ($\rho_{corr}$), respectively. $T_c$ is determined from the 50% resistivity drop from the residual resistivity, as shown in figures 2(a) and (b). The $T_c$ variations in irradiation are well expressed by one curve using $\rho_{corr}$.

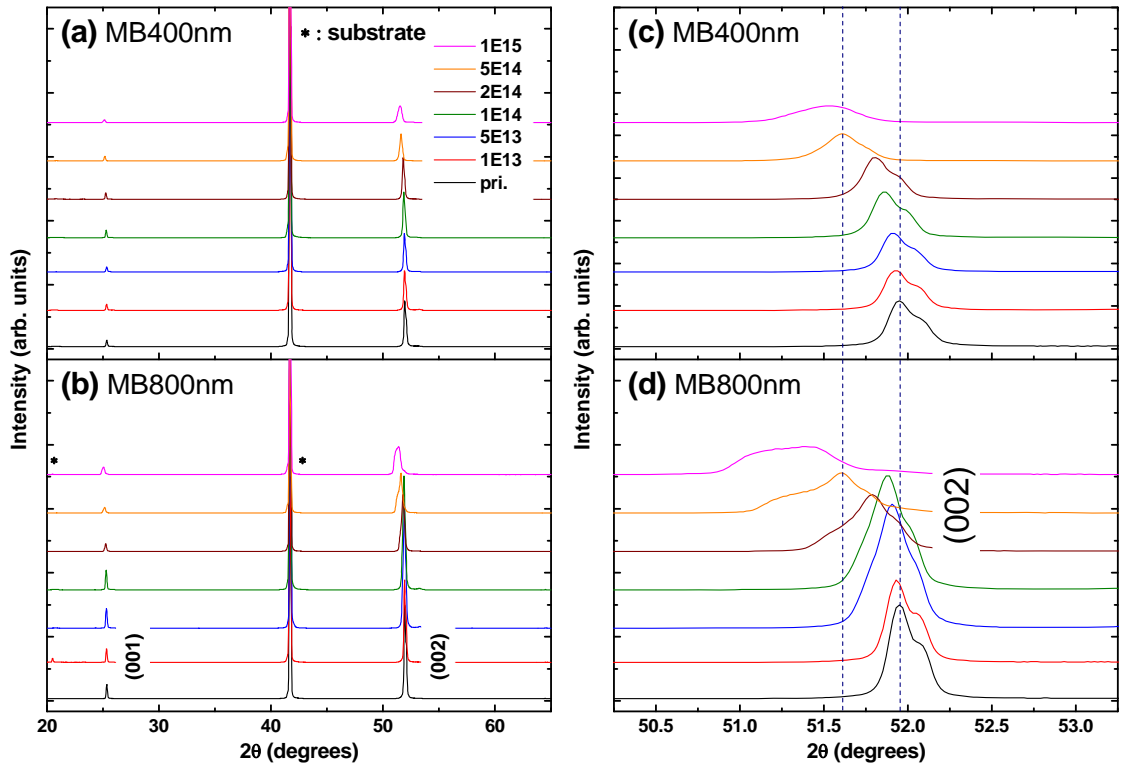

Figure 4. (a) and (b) X-ray diffraction (XRD) patterns of MB400nm and MB800nm, respectively, with various fluences. (c) and (d) Enlarged view near (002) peak for MB400nm and MB800nm, respectively. The shift in the peak position increases with increasing fluence, and the (002) peak of MB800nm becomes much broader at large fluences compared to that of MB400nm due to the implanted C ions inside the film. Dashed lines are guides for the eyes.

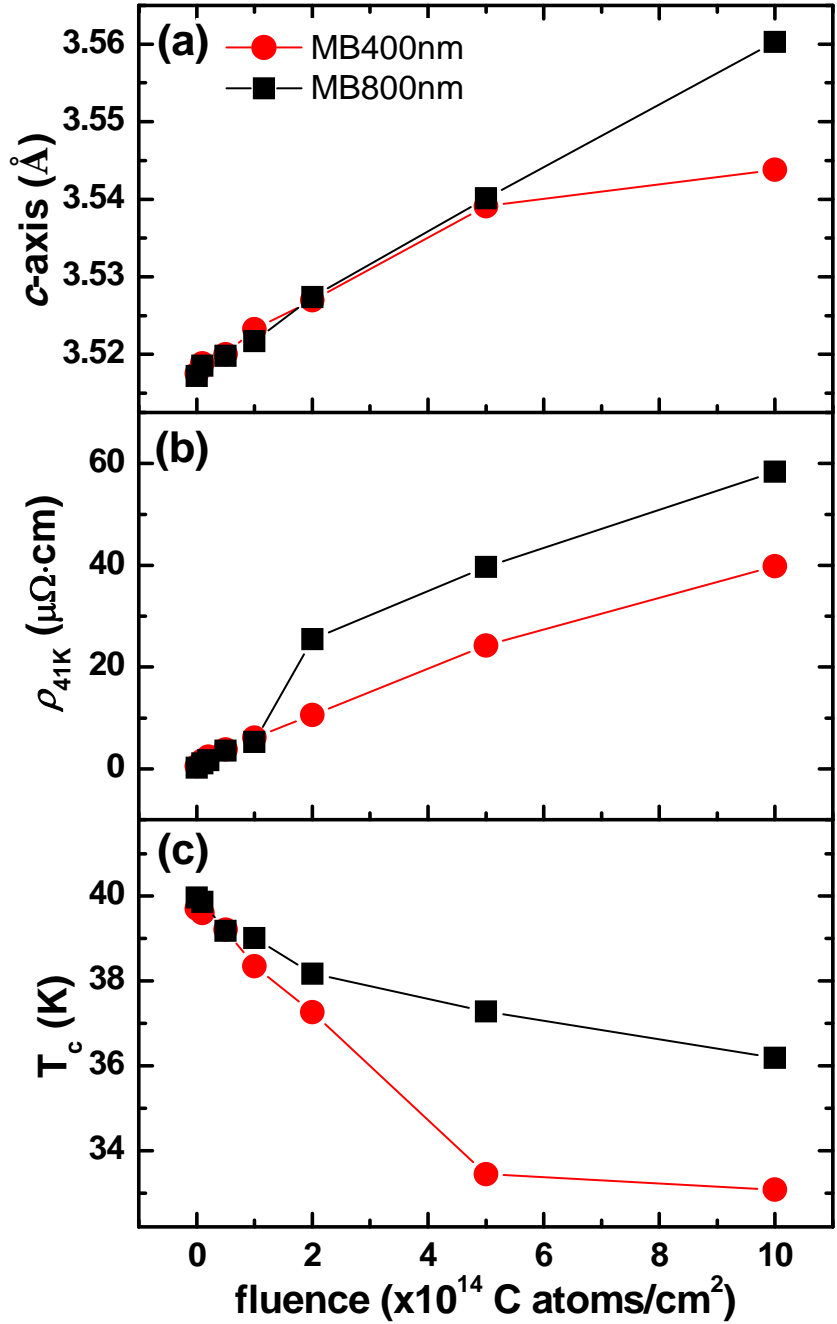

Figure 5. Fluence dependence of (a) $c$-axis lattice parameter, (b) residual resistivity ($\rho_{41K}$) and (c) $T_c$. $T_c$ of MB400nm decreased more rapidly than that of MB800nm with increasing fluence, on the other hand, $\rho_{41K}$ of MB800nm becomes much larger than that of MB400nm at the fluence of $2 \times 10^{14}$ C atoms/cm$^2$.

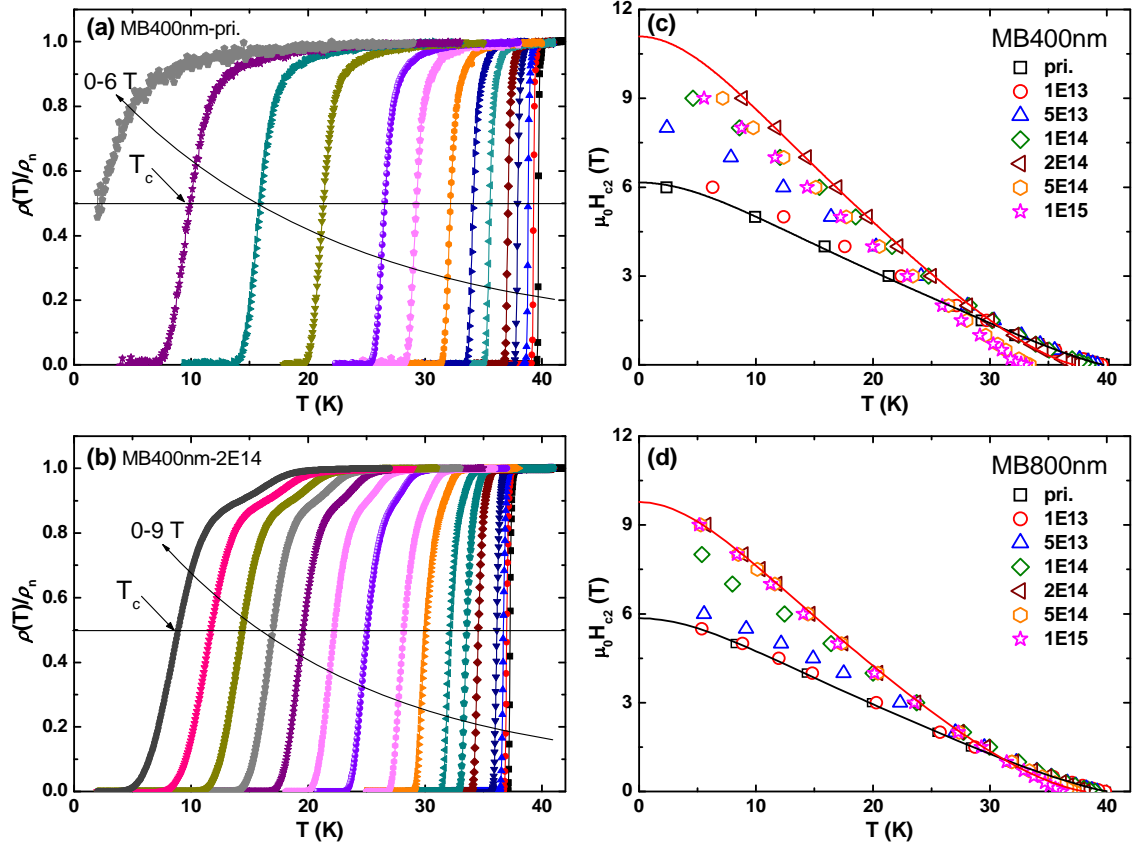

Figure 6. Temperature dependence of normalized resistivity, $\rho(T)/\rho_n$, for (a) pristine (pri.) and (b) $2 \times 10^{14}$ C atoms/cm$^2$ (2E14) C-ion irradiated MB400nm in magnetic fields. Upper critical field ($\mu_0 H_{c2}$) as a function of temperature for (c) MB400nm and (d) MB800nm for various fluences. Solid lines in figures 6(c) and (d) representatively show the fitting curves for the $H_{c2}(T)$ of the pristine and irradiated (2E14) MgB$_2$ films, obtained from two-band GL theory.

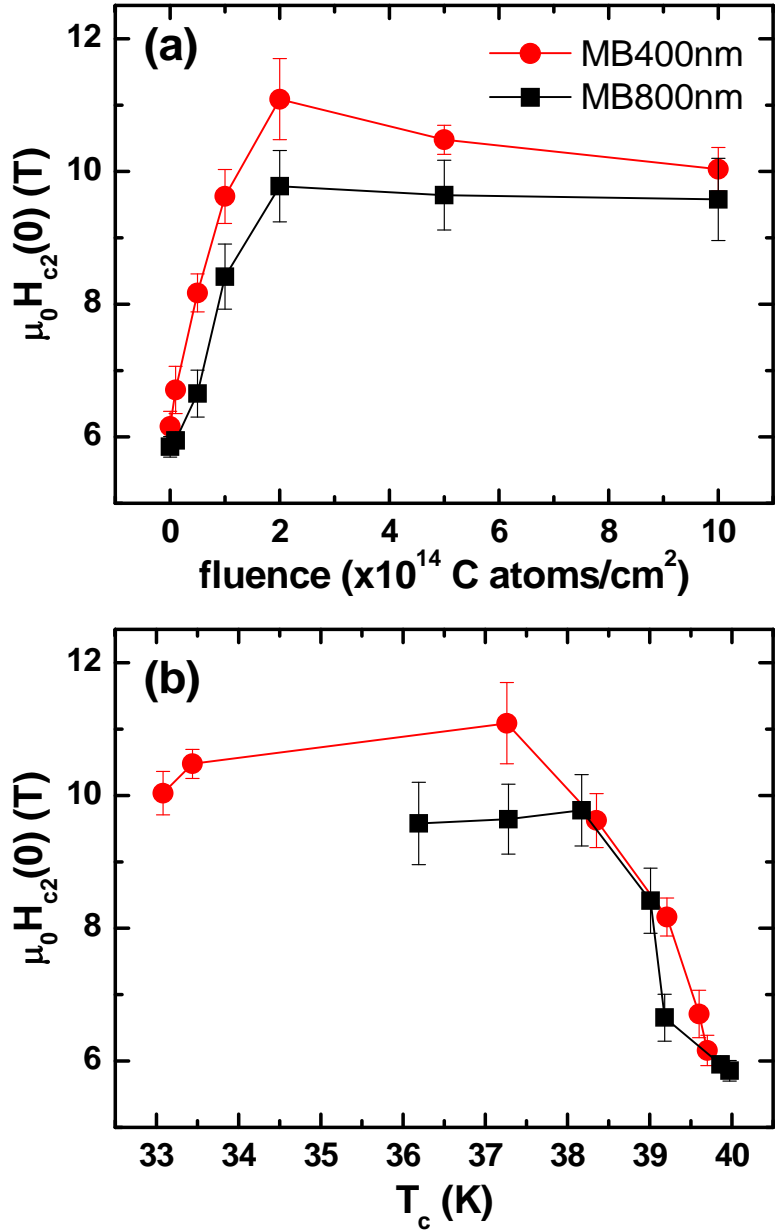

Figure 7. Zero-temperature upper critical fields, $\mu_0 H_{c2}(0)$, of MB400nm and MB800nm as functions of (a) fluence and (b) superconducting transition temperature ($T_c$). Both samples show the largest increase of $\mu_0 H_{c2}(0)$ at the fluence of $2 \times 10^{14}$ C atoms/cm$^2$. The error bars on the $\mu_0 H_{c2}(0)$ reflect uncertainties in the measured $H_{c2}(T)$s and fitting curves given by two-band GL theory.

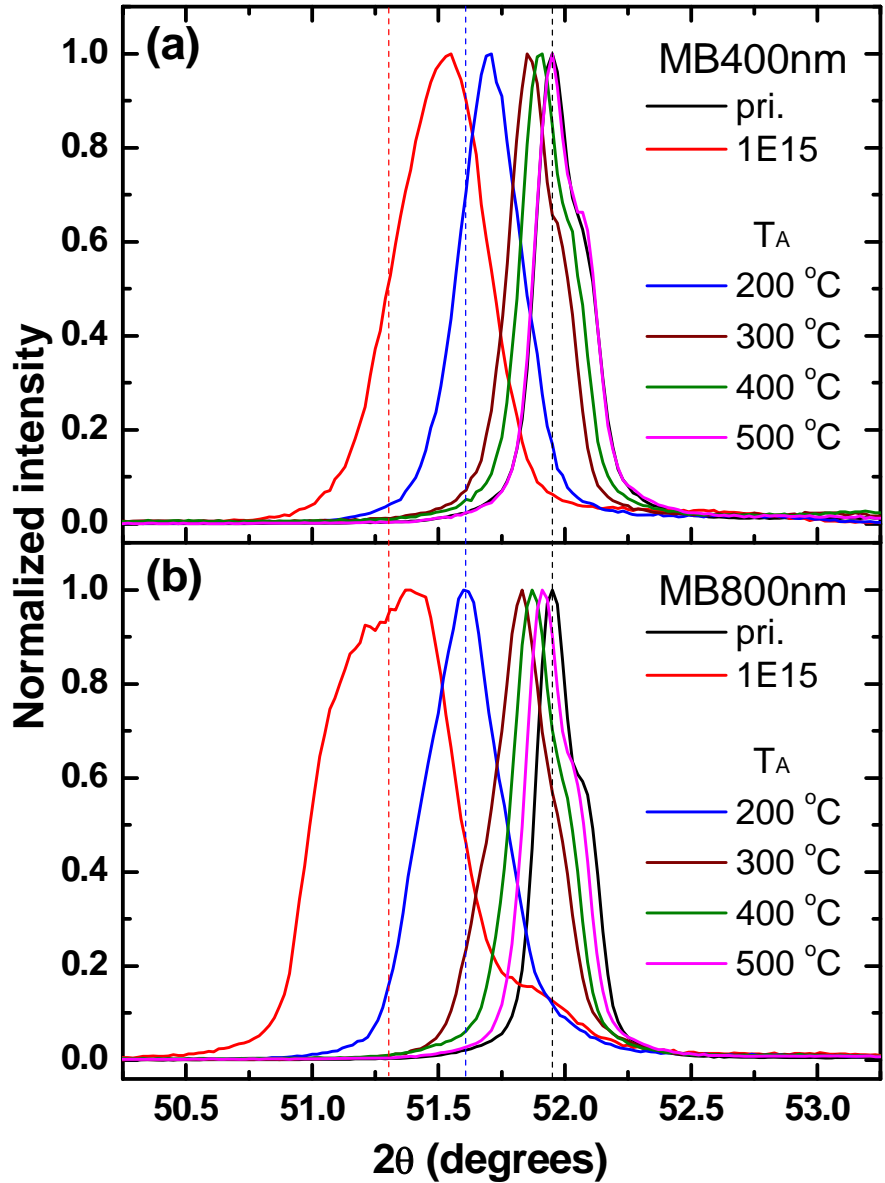

Figure 8. XRD of $\theta - 2\theta$ scans around (002) peak for (a) $1 \times 10^{15}$ C atoms/cm$^2$ (1E15) irradiated MB400nm and (b) MB800nm after thermal annealing at various temperatures. The moved peak position due to irradiation is gradually restored to its position in the pristine state with increasing annealing temperature ($T_A$), and the (002) peak position of MB400nm fully restored after annealing at 500 °C. The dashed lines are guides for the eyes for comparison of MB400nm and MB800nm.

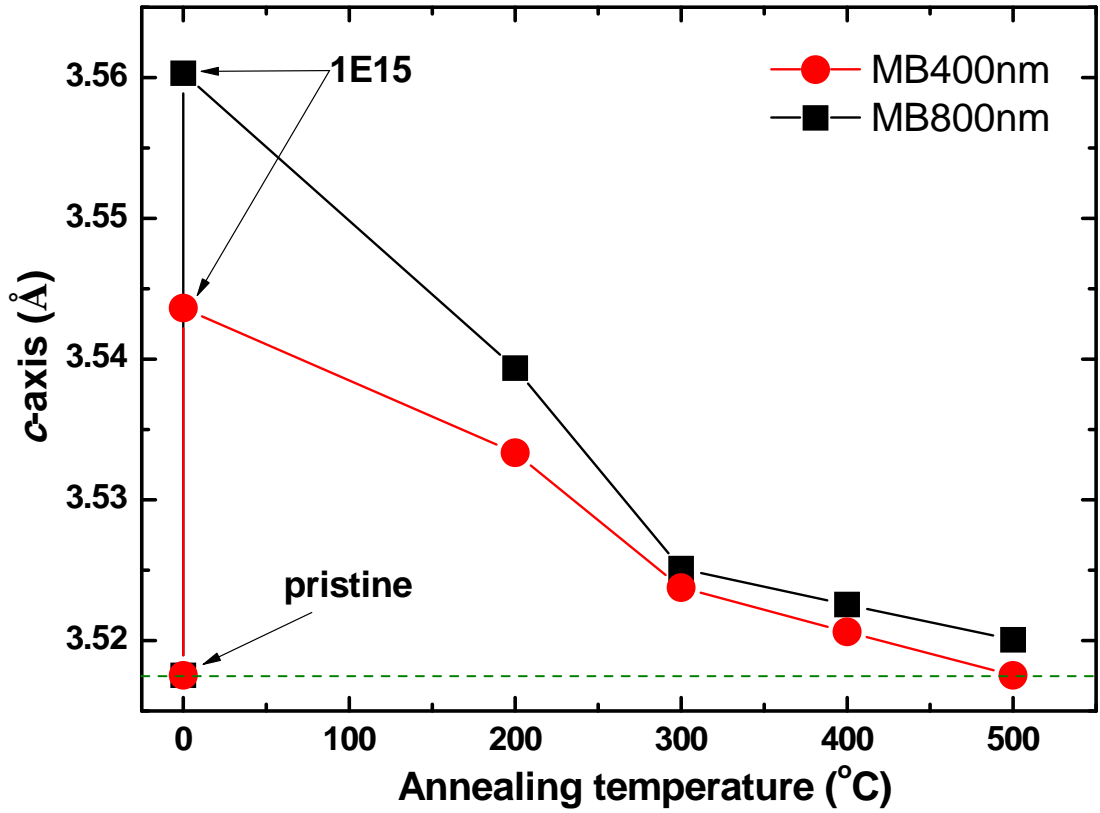

Figure 9. *c*-axis lattice parameter as a function of annealing temperature for $1 \times 10^{15}$ C atoms/cm$^2$ (1E15) irradiated MB400nm and MB800nm, calculated from the (002) peak shown in figure 8. The dashed line is a guide for the eyes.

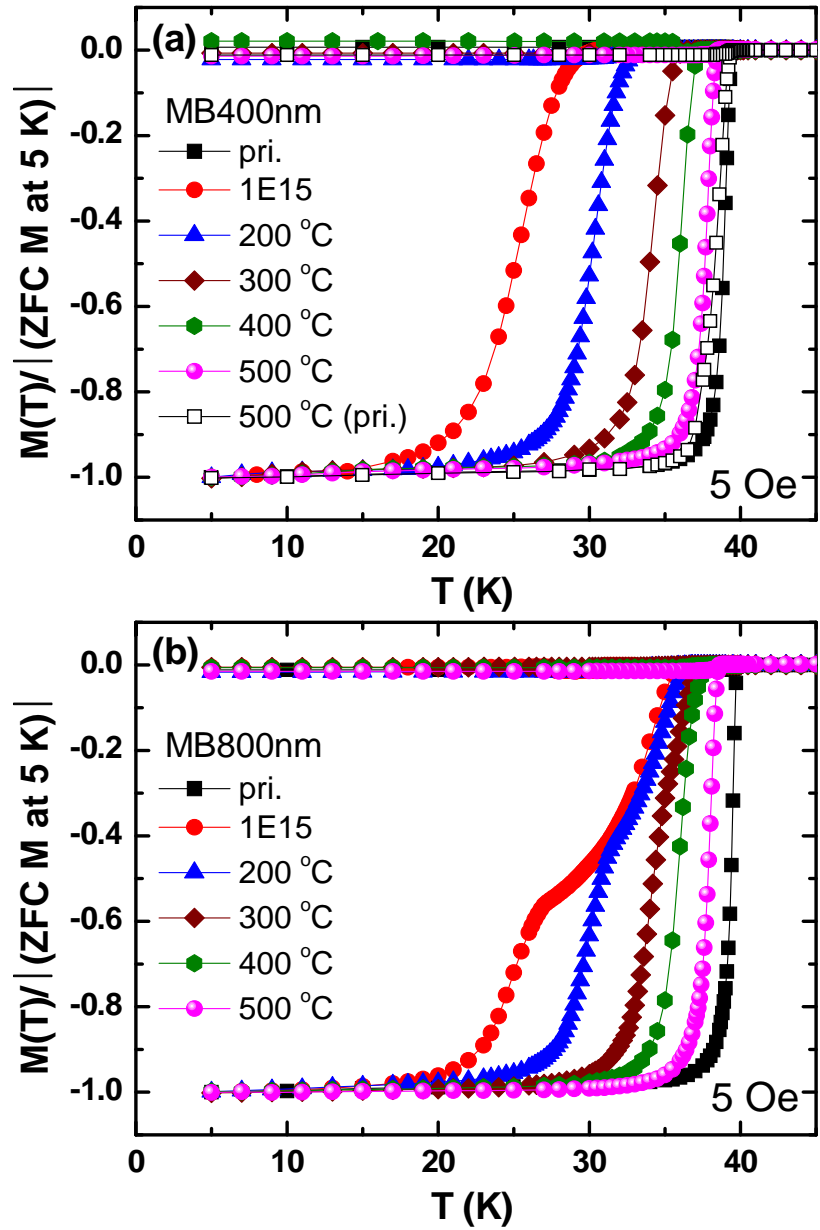

Figure 10. Temperature dependence of magnetization ($M$) for $1 \times 10^{15}$ C atoms/cm$^2$ (1E15) irradiated (a) MB400nm and (b) MB800nm after thermal annealing at various temperatures. $M(T)$ values are normalized by each ZFC $M$ value at 5 K for comparison. The superconductivity of both irradiated MB400nm and MB800nm films is almost restored to that in the pristine state after annealing at 500 °C.

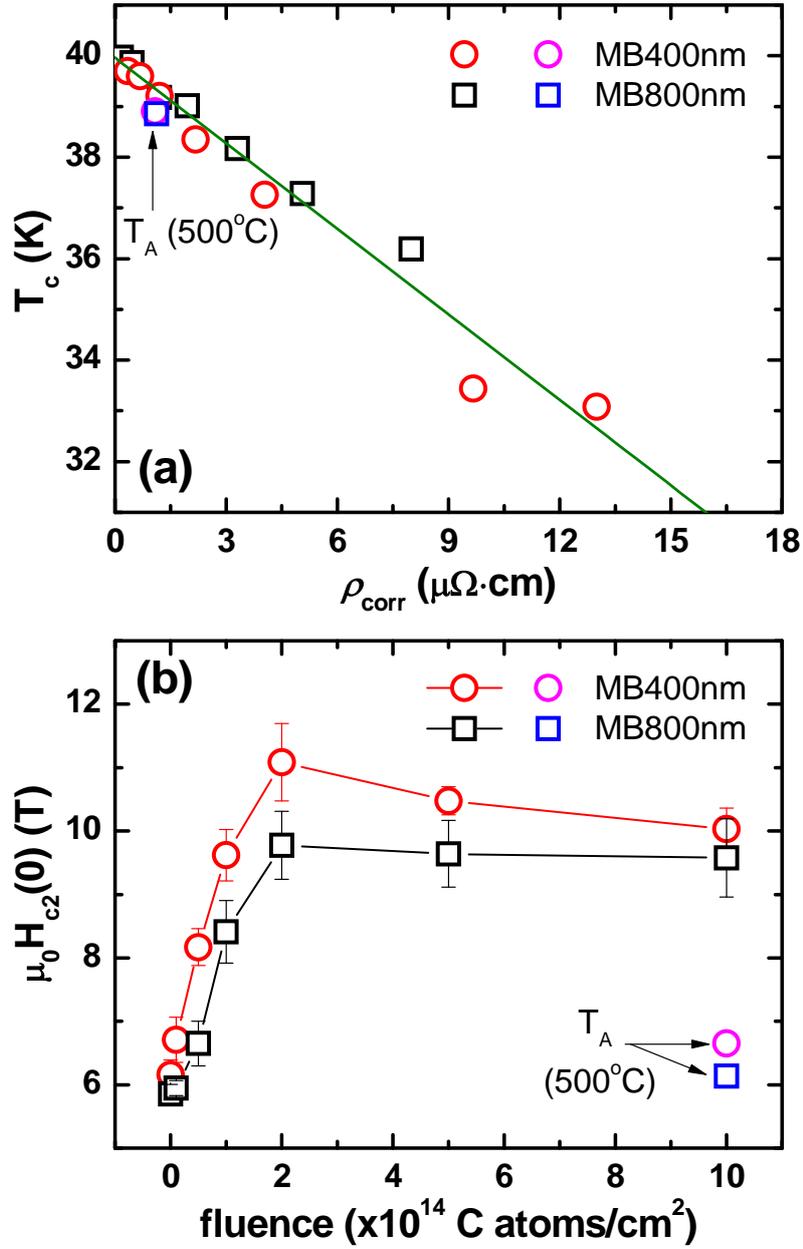

Figure 11. (a) $T_c$ for irradiated MB400nm and MB800nm and sequentially annealed $1 \times 10^{15}$ C atoms/cm$^2$ (1E15) irradiated films at $T_A = 500$ °C vs. corrected residual resistivity, $\rho_{corr}$. $T_c$ variation after annealing is also well expressed by the function of $\rho_{corr}$. (b) Change in $\mu_0 H_{c2}(0)$ for irradiated MB400nm (1E15) and MB800nm (1E15) after annealing at $T_A = 500$ °C. $\mu_0 H_{c2}(0)$ values for both films after annealing largely decrease and are close to those for the pristine films.

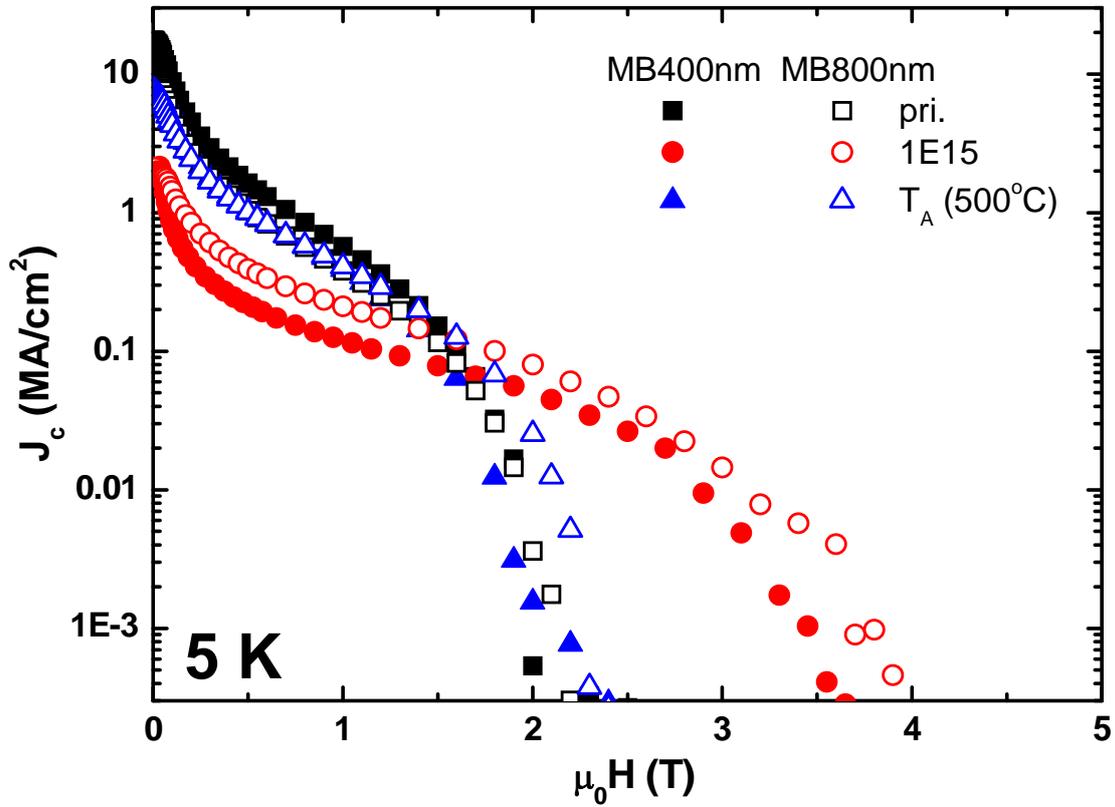

Figure 12. Critical current density ($J_c$) at 5 K as a function of magnetic field for MB400nm and MB800nm: pristine, $1 \times 10^{15}$ C atoms/cm$^2$ (1E15) irradiated, and annealed at 500 °C. Field performance of $J_c$ for irradiated films (1E15) became much stronger than that of pristine (pri.) samples despite a large suppression of low-field $J_c$. Interestingly, field dependence of $J_c$, $J_c(H)$, for both the irradiated films became comparable with $J_c(H)$ for pristine films after annealing them at $T_A$ = 500 °C.